\documentclass[pra,aps,reprint,a4paper,superscriptaddress,longbibliography,floatfix]{revtex4-1}
\usepackage{times}
\usepackage{epsfig}
\usepackage{amsfonts}
\usepackage{amsmath}
\usepackage{amssymb}
\usepackage{color}
\usepackage{multirow}
\usepackage[normalem]{ulem}
\usepackage{latexsym}
\usepackage{amsfonts}
\usepackage{mathrsfs}
\usepackage{natbib}
\usepackage{verbatim}

\usepackage[colorlinks=true,linkcolor=blue,citecolor=magenta,urlcolor=blue]{hyperref}

\def\endproof{\vrule height6pt width6pt depth0pt}

\begin{document}


\title{Quantum predictions for an unmeasured system cannot be simulated with a finite-memory classical system}



\author{Armin~Tavakoli}
\email{armin.tavakoli@unige.ch}
\affiliation{Groupe de Physique Appliqu\'ee, Universit\'e de Gen\`eve, CH-1211 Gen\`eve, Switzerland}
\affiliation{Department of Physics, Stockholm University, S-10691 Stockholm, Sweden}

\author{Ad\'an~Cabello}
\email{adan@us.es}
\affiliation{Departamento de F\'{\i}sica Aplicada II,
Universidad de Sevilla, E-41012 Sevilla, Spain}


\begin{abstract}
We consider an ideal experiment in which unlimited nonprojective quantum measurements are sequentially performed on a system that is initially entangled with a distant one. At each step of the sequence, the measurements are randomly chosen between two. However, regardless of which measurement is chosen or which outcome is obtained, the quantum state of the pair always remains entangled. We show that the classical simulation of the reduced state of the distant system requires not only unlimited rounds of communication, but also that the distant system has infinite memory. Otherwise, a thermodynamical argument predicts heating at a distance. Our proposal can be used for experimentally ruling out nonlocal finite-memory classical models of quantum theory.
\end{abstract}


\maketitle


\section{Introduction}


It has been shown recently that an experiment in which a single quantum system is subjected to many sequential measurements, each randomly chosen among $n$ alternatives, cannot be simulated with a classical system that has access only to finite memory. Such a classical simulation has an additional cost, namely, that after sufficiently many measurements the system dissipates heat. Specifically, the amount of heat per measurement tends to infinity and the divergence is linear in $n$ \cite{CGGLW16}. 

Such finite-memory classical simulations can be put in one-to-one correspondence with interpretations of quantum theory in which measurement outcomes are governed by intrinsic properties and, in addition, satisfy some assumptions \cite{CGGLW16}. Therefore, an experiment testing the presence or absence of such heat would rule out some interpretations of quantum theory. However, such an experiment is exposed to the practical problem of the implementation of the sequential measurements themselves producing heat, making it difficult to distinguish the hypothetical heat emitted by the finite-memory classical systems. Therefore, an interesting problem is whether a similar phenomenon can be demonstrated using sequential measurement that are {\em not} performed on the same physical system as from which the heat would originate. Such an experiment would require at least two systems, one that is being repeatedly measured and one in which the heat could appear. 

In order for measurements on one system to influence the quantum state of the other, the joint state of the two systems must have some entanglement. A complete projective local measurement performed on an entangled state renders the postmeasurement state separable. Thus, a second local measurement can no longer change the quantum state of the distant system. Therefore, the local measurements must necessarily be nonprojective positive-operator-valued measures (POVMs) in order to both induce a change in the local state of the distant system and retain this ability in a subsequent measurement. Sequential nonprojective local measurements have previously been shown useful in Bell experiments \cite{SGGP15} and random number generation \cite{CJAHWA17}. Here we show that sequential measurements can also be used to distinguish the predictions of quantum theory from classical simulations with finite memory in experiments involving two distant entangled systems. 

In Sec.\ \ref{protocol}, we introduce a protocol in which entanglement is preserved indefinitely for all measurement choices. Then, in Sec.\ \ref{cost}, we compute the cost of classically simulating some possible predictions of quantum theory for this experiment. We show that, in addition to an always increasing (but finite) amount of communication required for simulating entanglement in standard Bell experiments \cite{BCT99,Pironio03}, a thermodynamical analysis imposes an additional and qualitatively different cost: infinite local memory. Otherwise, an experiment would be able to detect the heat emitted by the system that is not measured.


\begin{figure*}[t]
	\vspace{-2.4cm}
	\includegraphics[width=1 \textwidth]{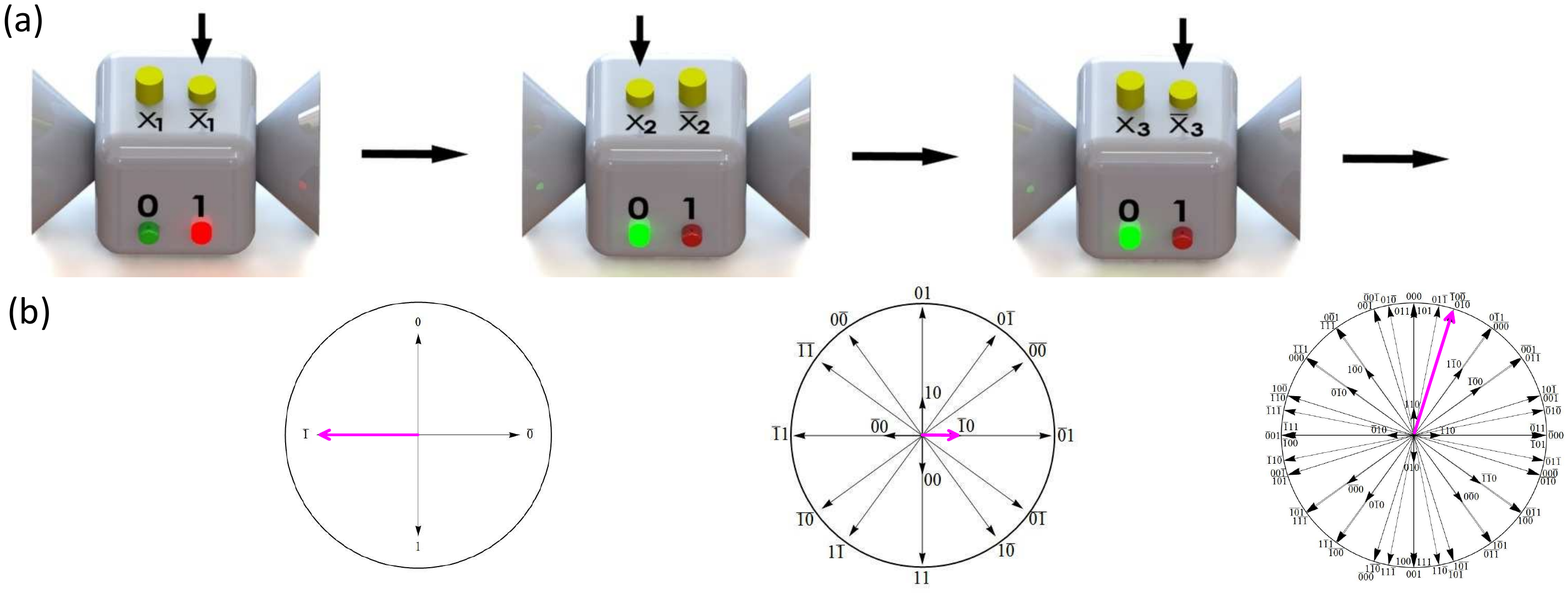}
	\vspace{-4.0cm}
	\caption{\label{Fig1} (a) From left to right, Alice's sequential measurements at times $t_1 < t_2 < t_3$, respectively. At each $t_k$, Alice performs a measurement, either $x_k$ or $\bar{x}_k$. Each measurement has two possible outcomes: $0$ and $1$. Alice's measurements are such that the state of the two qubits after her measurement is always entangled. (b) From left to right, possible reduced states of Bob's qubit after Alice's measurements at $t_1 < t_2 < t_3$, respectively. States are represented by nonunit arrows in the equatorial plane of the Bloch sphere. For example, $\bar{1}$ denotes the state when Alice measured $\bar{x}_1$ at $t_1$ and obtained outcome $1$; $\bar{1}0$ denotes the state when Alice measured $\bar{x}_1$ at $t_1$ and $x_2$ at $t_2$ and obtained outcomes $1$ and $0$, respectively. Bob's states highlighted in purple are those produced in the particular sequence of Alice's measurements and outcomes shown in (a).}
\end{figure*}


\section{Protocol}
\label{protocol}


\subsection{Scenario}
\label{scenario}


We consider two parties, Alice and Bob, who at time $t_0$ share two qubits in a maximally entangled state 
\begin{equation}
\label{state0}
|\psi_0\rangle=\frac{1}{\sqrt{2}}\left(|00\rangle+|11\rangle\right).
\end{equation} 
At later times $t_1 < t_2 < \cdots < t_N$, Alice randomly chooses between two measurements $x_k$ and $\bar{x}_k$ and performs this measurement on her qubit. Each measurement has two possible outcomes denoted by $0$ and $1$. The two measurements between which Alice measures are not preestablished, but depend on the previous measurements and outcomes. Bob does not perform any operation over the course of the protocol. However, at any time, the parties can stop the protocol and perform measurements (including Bob) to test some predictions of quantum theory. 

Any of Alice's measurement at time $t_k$, denoted by $j_k\in\{x_k,\bar{x}_k\}$, will be a two-outcome POVM which has, associated with outcome $0$, the POVM element $E_k^{j_k}=K_{\hat{n}_{j_k}}(\mu_{k})K_{\hat{n}_{j_k}}(\mu_{k})^\dagger$, where $K_{\hat{n}_{j_k}}(\mu_{k})$ is the Kraus operator \cite{Kraus83}
\begin{multline}
\label{Kraus}
K_{\hat{n}_{j_k}}(\mu_{k})\!=\!\cos\left(\mu_{k}\right)|\hat{n}_{j_k}\rangle\langle \hat{n}_{j_k}|+\sin\left(\mu_{k}\right)|-\hat{n}_{j_k}\rangle\langle -\hat{n}_{j_k}|,
\end{multline}
where $\hat{n}_{j_k}$ is a vector on the Bloch sphere that will be specified later. This POVM is a noisy version of the measurement represented by a Pauli matrix along $\hat{n}_{j_k}$. The amount of noise is controlled by the value of $\mu_{k}\in[0,\pi/2]$ and will be specified later. If $\mu_{k}\in\{0,\pi/2\}$, the measurement is projective. If $\mu_{k}=\pi/4$, then $K_{\hat{n}_{j_k}}=\openone/2$, implying a noninteractive measurement. Other values of $\mu_{k}$ correspond to weak measurements.

In addition, we assume that the time evolution is trivial, that is, that the state of Alice's and Bob's qubits just after Alice's measurement at $t_{k}$ is the state just before Alice's measurement at $t_{k+1}$, and is determined by Alice's sequence of measurements and outcomes at $\{t_1,\ldots, t_k\}$. The list of measurements and outcomes of Alice from $t_1$ to $t_k$ will be denoted by $l_k$.


\subsection{Choosing sequential measurements that always enable Bell inequality violation}


One of the features of the protocol that we are about to introduce is that, at each $t_k$, for each pair of measurements of Alice, there exist two measurements that Bob {\em could} perform (if the parties agreed to stop the protocol at this particular $t_k$) such that the outcome statistics of Alice and Bob would violate the Clauser-Horne-Shimony-Holt (CHSH) inequality \cite{CHSH69}.
Recall that in a CHSH experiment, Alice and Bob perform measurements $A_i$ and $B_j$, respectively, with $i,j\in \{0,1\}$, on shared pairs of systems. The measurement on each system is chosen independently and randomly. Any local realistic model of the outcome statistics must satisfy the CHSH inequality
\begin{equation}\label{chsh2}
	S_{\text{CHSH}}\equiv \langle A_0B_0\rangle +\langle A_0B_1\rangle +\langle A_1B_0\rangle -\langle A_1B_1\rangle \leq 2,
\end{equation}
where $\langle \cdot\rangle$ denotes expectation value. 

The following lemma (which is a corollary of the main result of Ref.\ \cite{WPF09}) explains why it is possible to achieve the feature described above.


{\em Lemma.} Consider any pure entangled state $\lvert \Psi_\eta \rangle=\cos\left(\eta\right)|00\rangle+\sin\left(\eta\right)|11\rangle$, with $\eta \in (0,\pi/2)$. For every $\lvert \Psi_\eta\rangle$, Alice can find measurements associated with Kraus operators \eqref{Kraus} with $\hat{n}_{j_k}$ equal to $\left(0,0,1\right)$ and $\left(1,0,0\right)$, respectively (i.e., noisy measurements of $\sigma_z$ and $\sigma_x$), for which she can choose a noise parameter $\mu\notin \{0,\pi/2\}$ such that there exists two projective measurements for Bob leading to outcome statistics violating the CHSH inequality \eqref{chsh2}.


\textit{Proof.} The Bloch vectors associated with the measurements $A_0$ and $A_1$ of Alice are $\left[0,0,\cos\left(2\mu\right)\right]$ and $\left[\cos\left(2\mu\right),0,0\right]$, respectively. These are unnormalized for $\mu \notin \{0,\pi/2\}$. Let us choose the Bloch vectors representing Bob's measurements $B_0$ and $B_1$ to be of the form $\left[\cos\left(\theta\right),0,\sin\left(\theta\right)\right]$ and $\left[-\cos\left(\theta\right),0,\sin\left(\theta\right)\right]$, respectively, for some $\theta$. These Bloch vectors are normalized and hence correspond to projective measurements. A direct computation of $S_{\text{CHSH}}$ in \eqref{chsh2} gives
\begin{equation}\label{chsh}
S_{\text{CHSH}}=2\cos\left(2\mu\right)\left[\sin\left(\theta\right)+\sin\left(2\eta\right)\cos\left(\theta\right)\right].
\end{equation}
We choose $\theta$ so that $S_{\text{CHSH}}$ is maximal, i.e., we solve the equation $\partial S_{\text{CHSH}}/\partial \theta=0$. The solution of our interest is $\theta=\arctan\left[1/\sin\left(2 \eta\right)\right]$, which is independent of $\mu$. Inserting this in Eq.\ \eqref{chsh} we find
\begin{equation}\label{chshvalue}
S_{\text{CHSH}}=\sqrt{6-2\cos\left(4 \eta\right)}\cos\left(2\mu\right).
\end{equation}
The minimal value of the square root is $2$ and is achieved for product states. Therefore, for every entangled state corresponding to $\eta \notin \{0,\pi/2\}$, the square root is larger than $2$. Hence, if Alice chooses her noise parameter $\mu$ such that
\begin{equation}\label{parameter}
0<\mu<\frac{1}{2}\arccos\left[\frac{2}{\sqrt{6-2\cos\left(4\eta\right)}}\right] \equiv F(\eta),
\end{equation}
then the outcome statistics of Alice and Bob will violate the CHSH inequality \eqref{chsh2}.\hfill \endproof


\subsection{Protocol}


Let us now describe the protocol itself. 

(0) At time $t_0$, Alice and Bob share the maximally entangled state $\lvert \psi_0\rangle$ given in Eq.\ \eqref{state0}.

(1a) At $t_1 > t_0$, Alice chooses some nonzero $\mu_1<F(\pi/4)=\pi/8$. Then, she randomly chooses between $x_1$ and $\bar{x}_1$, each of them associated with a Bloch vector $(0,0,1)$ and $(1,0,0)$, respectively. Alice's choice is denoted by $j_1$. Then, Alice performs the measurement $\{E^{j_1}_1,\openone-E^{j_1}_1\}$. The Lemma enssures that, for the state before the measurement and Alice's two possible measurements, there are two possible measurements on Bob's system violating the CHSH inequality.

(1b) From her observed outcome, Alice calculates the postmeasurement state $|\psi^{l_1}_1\rangle$ of the two qubits. This state is necessarily pure and entangled, and can be written in the form
\begin{equation}\label{rot}
|\psi^{l_1}_1\rangle=U^{l_1}_A\otimes U^{l_1}_B \left[\cos\left(\theta^{l_1}\right)|00\rangle+\sin\left(\theta^{l_1}\right)|11\rangle\right],
\end{equation}
where $\theta^{l_1}\notin \{0,\pi/2\}$ and $U^{l_1}_A$ and $U^{l_1}_B$ are unitary operators. Here $\theta^{l_1}$ does not refer to an actual operation but is a hypothetical angle which would maximize Eq.\ (\ref{chshvalue}). Then Alice applies on her qubit the unitary $\left(U^{l_1}_A\right)^\dagger$, which cancels the unitary $U^{l_1}_A$ in Eq.\ \eqref{rot}. After Alice's actions at $t_1$, the reduced state of Bob's qubit is one of four possible states (see Fig.\ \ref{Fig1}).

(2a) At $t_2$, Alice again chooses some positive $\mu_2<F(\theta^{l_1})$. She makes a random choice of measurement $j_2\in\{x_2,\bar{x}_2\}$ associated with Bloch vectors $(0,0,1)$ and $(1,0,0)$, respectively, and performs the measurement $\{E^{j_2}_2,\openone-E^{j_2}_2\}$. Again, the Lemma ensures that, for the state before the measurement and Alice's two possible measurements, there are two possible measurements on Bob's system violating the CHSH inequality.

(2b) From her observed outcome, Alice calculates the new postmeasurement state $|\psi^{l_2}_2\rangle$ of the two qubits. Just as in (1b), Alice rotates her reduced state back to the computational basis by applying a suitable unitary. After Alice's actions at $t_2$, the reduced state of Bob's qubit is one of $16$ possible states (see Fig.\ \ref{Fig1}).

Alice continues this process of measuring, recording the outcome, and choosing the next measurement indefinitely. That is:

(ta) At $t_k$, she chooses some positive $\mu_k<F\left(\theta^{l_{k-1}}\right)$, randomly chooses a measurement $j_k\in\{x_k,\bar{x}_k\}$ associated with Bloch vectors $(0,0,1)$ and $(1,0,0)$, respectively, and performs the measurement $\{E^{j_k}_k,\openone-E^{j_k}_k\}$. The Lemma guarantees that, for the state before the measurement and Alice's two possible measurements, there are two possible measurements on Bob's system violating the CHSH inequality.

(tb) From her observed outcome, Alice calculates the postmeasurement state $|\psi^{l_k}_k\rangle$ which takes the form
\begin{equation}\label{rot2}
|\psi^{l_k}_k\rangle=U^{l_k}_A\otimes U^{l_k}_B \left[\cos\left(\theta^{l_k}\right)|00\rangle+\sin\left(\theta^{l_k}\right)|11\rangle\right],
\end{equation}
for some angle $\theta^{l_k}\notin \{0,\pi/2\}$. Subsequently, she undoes the rotation of her local state by applying $\left(U_A^{l_k}\right)^\dagger$. This renders the reduced state of Bob's qubit in one of $4^k$ possible states (see Fig.\ \ref{Fig1}).


\subsection{Properties of the protocol}


At each time $t_k$, Alice's alternative measurements are {\em both} nonprojective and depend on Alice's previous choices of measurements and also on the outcomes of the previous measurements. This way, the initial entanglement is never consumed regardless of Alice's performed measurements and observed outcomes and Alice's two measurement options enable a violation of the CHSH inequality.


\begin{table}[t]
	\centering 
	\begin{tabular}{c c c c c c c} 
		\hline\hline 
		$\substack{\text{Time}}$ & $\mu$ & $\substack{\text{Number of possible}\\ \text{states of Bob's qubit}}$ & $\substack{\text{Smallest}\\\text{negativity}}$ & $\substack{\text{Largest}\\\text{negativity}}$ & $\substack{\text{Smallest}\\ \text{value of }S_{\text{CHSH}}}$ & $\substack{\text{Largest}\\ \text{value of }S_{\text{CHSH}}}$ \\ [0.5ex] 
		\hline 
		$t_0$ & & 1 & $0.5$ & $0.5 $ & & \\
		$t_1$ & $\pi/9$ & 4 & $0.3214$ & $0.3214$ & $2.1667$ & $2.1667$ \\ 
		$t_2$ & $\pi/12$ & 16 & $0.0966$ & $0.4774$ & $2.0590$ & $2.0590$\\
		$t_3$ & $\pi/40$ & 64 & $0.0077$ & $0.4887$ & $2.0119$ & $2.7313$\\
		$t_4$ & $\pi/500$ & 256 & $0.00005$ & $0.4902$ & $2.00008$ & $2.7965$ \\
		\hline 
		\hline
	\end{tabular}
	\caption{Data from the first four time steps in one possible execution of the quantum protocol: choices of the noise parameter for Alice's measurement at $t_k$, the number of different local states of Bob just after $t_k$, the smallest and largest negativity of the $4^k$ possible global states just after $t_k$, and the smallest and largest values of $S_{\text{CHSH}}$ achieved with the $4^{k-1}$ possible states. The choices of $\mu$ carry no special significance other than that they satisfy the relation $0<\mu_k<F(\theta^{l_{k-1}})$ for all $k$.}
	\label{Tab1} 
\end{table}


To illustrate the properties of the protocol, in Table \ref{Tab1} we display data from the first few steps of one possible execution of the protocol. There we can see that at each time step, the measurement of Alice becomes stronger without ever becoming projective. Furthermore, the entanglement, quantified by the negativity \cite{ZHSL98}, remains nonzero. From $t_2$ onward, not all of the $4^k$ possible states just after $t_k$ contain the same amount of entanglement, and therefore we must consider the weakest possible entanglement. As displayed in Table \ref{Tab1}, the negativity of the weakest entangled state quickly decreases. However, some entanglement is always present. When choosing her noise parameter $\mu_k$, Alice ensures that even the weakest entangled state violates the CHSH inequality \eqref{chsh2}. That this is indeed the case can be seen from the corresponding smallest values of $S_{\text{CHSH}}$ in Table \ref{Tab1}. In contrast, from the largest possible negativity we see that the protocol sometimes, albeit with small probability, acts as a probabilistic entanglement amplification scheme \cite{OAN12,BBPS96}.


\section{Classical simulation of Bob's local state}
\label{cost}


\subsection{Cost 1: Unlimited rounds of communication}
\label{communication}


We now consider the cost of classically simulating the evolution of the quantum reduced state of Bob's system induced by Alice's sequential measurements on her distant system. A classical simulation of Bob's local state must be able to, at any time $t_k$, account for the statistical outcomes of every possible quantum measurement that Bob may apply. We will not consider this problem in its full generality, as it is not the emphasis of our work. For our purposes, it suffices to show that after each of Alice's sequential measurements, the classical simulation needs to be supplemented with some amount of communication. 

Since the postmeasurement state just after $t_k$ and the two measurement options Alice has at time $t_{k+1}$ could always be used, in conjunction with suitable measurements of Bob, to violate the CHSH inequality, any local realist model aiming to simulate these quantum predictions has to be supplemented with some communication \cite{BCT99,Pironio03}. Therefore, there must be a round of communication between Alice and Bob after every measurement performed by Alice. In this round, depending on her measurement and the resulting local realistic state of her system, Alice communicates some information to Bob. Communication from Bob to Alice is of no use since Bob does not perform any operations on his qubit over the course of the protocol. 

The critical observation is that the communication required to simulate the quantum predictions just after $t_{k}$ will not be enough to reproduce the quantum predictions after $t_{k+1}$. The reason is that, at $t_{k+1}$, the new quantum predictions could again be use to violate the CHSH inequality. Thus, regardless of Alice's measurement choice and observed outcome, any simulation of the predictions of quantum theory for the experiment based on a local realistic model complemented with communication requires unlimited rounds of communication. 

Note that the \textit{total} amount of communication required to simulate the ability of the state to violate the CHSH inequality at each time step, is finite. To show this, we use that the average amount of communication $\mathcal{C}$ required to simulate a nonsignaling probability distribution achieving the value $S_{\text{CHSH}}$ is given by $\mathcal{C}=S_{\text{CHSH}}/2-1$ \cite{Pironio03}. Let us denote the average communication over all possible postmeasurement states at time $t_k$ by $\mathcal{C}_k$. The total amount of communication is finite if $\bar{\mathcal{C}}\equiv \sum_{k=1}^{\infty}\mathcal{C}_k$ is finite. To show that $\bar{\mathcal{C}}$ is finite, we consider the states $|\psi_\eta\rangle$ which are unitarily equivalent to the states shared by Alice and Bob. Applying the Horodecki criterion \cite{HHH95} to $|\psi_\eta\rangle$, we find the maximal value of $S_{\text{CHSH}}$ at time $t_k$. It is a straightforward calculation to show that this quantity is an upper bound on the sum of the CHSH value \eqref{chshvalue} of $|\psi_\eta\rangle$, as obtained when applying a noisy measurement in our protocol at time $t_k$, and the average maximal CHSH value, obtained from applying the Horodecki criterion to the four possible postmeasurement states at $t_{k+1}$ weighted by the respective probability of obtaining each state. This argument can be repeated throughout the protocol and consequently $\bar{\mathcal{C}}<\sqrt{2}-1$, which is the communication cost of simulating a maximal violation of the CHSH inequality achieved with $\lvert \psi_0\rangle$.


\subsection{Cost 2: Unbounded local memory}
\label{memory}


At each time step in the protocol, Alice chooses with uniform probability between two measurement options. After each measurement of Alice, the number of possible reduced states of Bob's qubit quadruples. Any classical simulation must account for this exponentially increasing number of possible states. Since each of Alice's measurement choices is random, any classical simulation requires having at least the same number of local realist states as the number of pure quantum states achieved during the experiment. The proof is as follows.

A stochastic process is a one-dimensional chain of discrete random variables that attains values in a finite or countably infinite alphabet. An input-output process \cite{BC14} is a collection of stochastic processes in which each such process corresponds to all possible output sequences given a particular infinite input sequence. The experiment is an example of an input-output process. It has input alphabet $\{x_k,\bar{x}_k\}$ and output alphabet $\{0,1\}$. As shown in Ref.\ \cite{BC14}, for any input-output process there is a unique finite-state machine, i.e., an abstract machine that can be in exactly one of a finite number of states at any given time, with the following property: It has minimal entropy over the state probability distribution and maximal mutual information with the future output of the process given the past choices of inputs and past observed outputs, and the future input of the process. This machine is called the $\varepsilon$ transducer \cite{BC14} of the input-output process. It consists of the input and output alphabets, a set of causal states, and the set of conditional transition probabilities between the causal states. Each causal state is associated with the set of input-output pasts producing the same probabilities for all possible input-output futures. Thus, the causal states constitute equivalence classes for the set of input-output pasts. A causal state stores all the information about the past needed to predict the future output, but as little as possible of the remaining information overhead contained in the past. The Shannon entropy over the stationary distribution of the causal states represents the minimum internal entropy needed to be stored to optimally compute future outputs. It depends on how Alice's measurements are chosen; here we have assumed that they are selected from a uniform probability distribution with entropy one bit at each time step.
	
The number of causal states of the $\varepsilon$ transducer corresponding to our experiment is infinite. This implies that the classical system that simulates the experiment has to store new information in its memory. This leads to two possibilities: Either the memory is infinite and additional information can always be stored without needing to erase previous information, or the memory is finite and the system has to erase a part of it to allocate new information. However, due to Landauer's principle \cite{Landauer61}, the erasure of information has a thermodynamical cost. Landauer's principle states that the erasure of information in an information-carrying degree of freedom is accompanied by an associated increase of entropy in some non-information.carrying degree of freedom. 
There is strong evidence supporting the validity of Landauer's principle in both the classical and quantum domains \cite{P00,HSAL11,BA12,JGB14,RW14,LM15,PS16}. Since we are assuming a local realistic model supplemented by communication, the memory should be allocated in the local systems. Since Bob's quantum state and its classical counterpart (represented by a causal state of the $\varepsilon$ transducer) are changing after each of Alice's measurements, this implies that there should be some information erasure in the local memory associated with Bob's system. Therefore, after sufficiently many measurements of Alice, Bob's system begins to emit heat. Such heating at a distance is a form of signaling. 


\section{Conclusions}


We have introduced a protocol in which sequential nonprojective measurements are performed on one of two entangled systems while no measurements are performed on the other distant system. Regardless of which local measurements are chosen and which outcomes are obtained, both entanglement and the possibility of violating a Bell inequality {\em never} vanish. We showed that, to simulate the predictions of quantum theory for the local state of the distant system, it is not sufficient to supplement finite-memory classical models with unlimited rounds of communication. In addition, the distant system must have infinite memory. Whenever the distant system fails to have infinite memory, a thermodynamical argument implies that it will be heated at a distance after sufficiently many local measurements on its companion. 

Our protocol shows that (i) there is a way for experimentally ruling out nonlocal finite-memory classical models without measuring the system that will, hypothetically, emit heat and (ii) there are problems whose solution would require classical systems with infinite memory and communication but which can be solved combining sequential quantum measurements and entanglement.


\section*{Acknowledgements}


We thank Nicolas Brunner, Nicolas Gisin, Mile Gu, and Matthias Kleinmann for their comments on the manuscript, Gustavo Ca\~{n}as for his help with Fig.\ \ref{Fig1}, Jim Crutchfield for discussions, and Matthias Kleinmann for checking the calculations. This work was supported by the project ``Photonic Quantum Information'' (Knut and Alice Wallenberg Foundation, Sweden), Project No.\ FIS2014-60843-P, ``Advanced Quantum Information'' (MINECO, Spain), with FEDER funds, and the FQXi Large Grant ``The Observer Observed: A Bayesian Route to the Reconstruction of Quantum Theory.'' A.T.\ acknowledges financial support from the Swiss National Science Foundation (Starting Grant DIAQ).




\end{document}